\documentclass[prd,amsmath,amssymb,preprintnumbers,showpacs,showkeys]{revtex4}
\newcommand{\ip}[2]{{\langle #1\mid #2\rangle}}

\usepackage{amsmath,amssymb,mathrsfs}

\begin{document}
\title{Comments on ``Counter example to the quantum inequality''}
\author{Christopher J. Fewster}
\email{cjf3@york.ac.uk}
\affiliation{Department of Mathematics, University of York, Heslington, York, YO10 5DD, UK}
\date{\today}
\begin{abstract} 
In a recent preprint, Krasnikov has claimed that to show
that quantum energy inequalities (QEIs) are violated in curved spacetime situations,
by considering the example of a free massless scalar field in
two-dimensional de Sitter space. We show that this claim is incorrect, and based
on misunderstandings of the nature of QEIs. 
We also prove, in general two-dimensional spacetimes, 
that flat spacetime QEIs give a good approximation to the curved
spacetime results on sampling timescales short in comparison with
natural geometric scales.
\end{abstract}
\maketitle
%

%\section{}

Classically, the massless (minimally coupled) free scalar field obeys the weak energy
condition: it displays a nonnegative energy density to all observers
at all points in spacetime. Its quantised sibling is quite different, however,
admitting unboundedly negative energy densities at individual spacetime
points. Violations of the energy conditions are cause for concern, and
a considerable effort has been expended, beginning with the work of
Ford~\cite{Ford78}, in trying to understand what
constraints quantum field theory might place on such effects. 
It turns out that averages of the energy density along, for example, timelike
curves obey state-independent lower bounds called quantum inequalities,
or quantum energy inequalities (QEIs). 
In two-dimensional Minkowski space, for example, the massless free field obeys the
QEI bound~\cite{Flanagan97}
\begin{equation}
\int_\gamma \langle T_{ab}(\gamma(\tau))\rangle_\omega u^a u^b G(\tau) d\tau \ge
- \frac{1}{24\pi}\int_{-\infty}^\infty \frac{G'(\tau)^2}{G(\tau)} d\tau
\label{eq:flatQI}
\end{equation}
for all Hadamard states $\omega$,
where $\gamma$ is the worldline of an inertial observer parametrised by
proper time $\tau$ with two-velocity $u^a$, and $G$ is any smooth,
nonnegative sampling function of compact support [i.e., vanishing
outside a compact interval]. The right-hand side is large and negative
if $G$ is tightly peaked, but small if it is broadly spread. Thus
the magnitude and duration are constrained by a relationship reminiscent of
the uncertainty relations: in $d$-dimensional Minkowski space, 
the energy can density drop below $\rho_0<0$ for a time $\tau_0$ only if
$|\rho_0|\tau_0^d<\kappa_d$ for some (small) constant $\kappa_d$ ($\kappa_2=\pi/6$, for example). 

Many exotic spacetimes (wormholes, warp drives, etc) entail violations
of the weak energy condition and it has often been suggested that
quantum fields might provide the necessary distributions of
stress-energy. Quantum energy
inequalities provide a quantitative check on such proposals and have
been used to argue that exotic spacetimes are tightly constrained~\cite{FRworm,FPwarp}.
As no curved spacetime QEIs were available when these references were
written, they made use of flat spacetime
QEIs, and the validity of their conclusions depends on the assumption that quantum fields
in curved spacetimes are subject to the same
restrictions as those in flat spacetimes, at least on sampling
timescales short in comparison with natural geometric scales. We will
refer to this as the `usual assumption'. 

In a recent preprint~\cite{Kras04}, Krasnikov has claimed that the
usual assumption fails, on the basis of an explicit example in two-dimensional
de Sitter space. If true, this would necessitate a reevaluation of the
constraints obtained in~\cite{FRworm,FPwarp}. However, we will show here
that Krasnikov's claim is incorrect; moreover, we will prove that the
usual assumption is justified in all two-dimensional globally hyperbolic
spacetimes.  

Before doing so, let us consider the status of the usual assumption in general.
Contrary to what is claimed in~\cite{Kras04}, QEIs have been
established in curved spacetimes. Indeed, there are results for the free
scalar~\cite{FordPfenning98,FTeo,AGWQI,Flanagan02}, Dirac~\cite{Vollick00,FVdirac}, Maxwell and Proca
fields~\cite{Pfenning_em,FewsterPfenning} in various levels of
generality, including very general results. It is true that, with the exception
of~\cite{Vollick00,Flanagan02}, these bounds have been ``difference'' QEIs:
namely, the quantity bounded is the difference between the energy
density in the state of interest and that in a reference state. 
For example, in~\cite{AGWQI} a QEI was obtained for 
the scalar field in an arbitrary globally hyperbolic spacetime
$(M,g_{ab})$ for sampling along any smooth timelike curve $\gamma$,
which took the form
\begin{equation}
\int_\gamma \left[\langle T_{ab}(\gamma(\tau))\rangle_\omega 
-T_{ab}(\gamma(\tau))\rangle_{\omega_0}\right]  u^a u^b G(\tau)
d\tau \ge -{\cal Q}[M,g_{ab},\gamma,\omega_0,G]
\end{equation}
where $\omega_0$ is an (arbitrary, but fixed) reference Hadamard state.
This bound holds for arbitrary Hadamard states $\omega$ and any $G$ of the
form $G(\tau)=g(\tau)^2$ with $g$ real-valued, smooth
and compactly supported; an explicit formula for ${\cal Q}$ can be
given~\cite{AGWQI}. Now it is easy to see that an ``absolute'' QEI
follows immediately, simply by
correcting the ``difference'' bound by the renormalised energy density of the
reference state:
\begin{equation}
\int_\gamma \langle T_{ab}(\gamma(\tau))\rangle_\omega 
u^a u^b G(\tau)
d\tau \ge -{\cal Q}[M,g_{ab},\gamma,\omega_0,G]+
\int_\gamma \langle T_{ab}(\gamma(\tau))\rangle_{\omega_0}u^a u^b G(\tau)\,d\tau\,.
\end{equation}
Of course, the problem is that these expressions depend on the
reference state, and in a general spacetime it is not usually possible to write down a closed
form expression for the stress-tensor of any particular Hadamard state. But now
replace $G$ by its scaled version 
\begin{equation}
G_{\tau_0}(\tau)=\tau_0^{-1}G(\tau/\tau_0)\,,
\label{eq:scaled}
\end{equation}
which has the same integral as $G$ (for convenience we will take this to
be unity). The difference QEI bound is expected to scale as 
$\tau_0^{-d}$ in $d$-dimensions, and to approach the corresponding Minkowski space bound
for sufficiently small $\tau_0$. This indeed occurs in examples~\cite{FTeo}
and a general proof is probably not too difficult. On the other hand,
the second term will approach the constant value 
$\langle T_{ab}(\gamma(0))\rangle_{\omega_0}u^a u^b$ as $\tau_0\to 0$ and is therefore
swamped by the first term when $\tau_0$ is small enough. 

To establish the usual assumption we must quantify
how small is `small enough'. In examples, the difference QEI approaches
the corresponding Minkowski results on timescales short in comparison
with geometric scales, but there remains the problem of the constant
term arising from the reference state. It has not (yet) been ruled out that the
reference state could make an anomalously large
contribution~\footnote{In order to invalidate the usual assumption,
however, it would be necessary to show that {\em all} Hadamard reference
states gave such an anomalous contribution.} in which case the timescale
$\tau_0$ might have to be chosen very much shorter than natural
geometric scales. In this case the QEI bound would be very weak, and
perhaps insufficient to constrain the geometry as in~\cite{FRworm,FPwarp}. At present there is,
therefore, a small gap in justifying the usual assumption in dimensions
greater than two, although there is a strong expectation that it can
be bridged. 

For massless fields in two dimensions, however, the situation is rather
different. Conformal invariance makes it possible to obtain explicit formulae for
``absolute'' QEI bounds, a fact first realised by
Vollick~\cite{Vollick00} and further developed by
Flanagan~\cite{Flanagan02}. The trick is to make a conformal
transformation back to two-dimensional Minkowski space, where the results
of~\cite{Flanagan97} can be employed. As we will now show, these bounds
permit us to prove the usual assumption in this setting. 

Consider the free massless scalar field on an arbitrary two-dimensional globally hyperbolic Lorentzian
spacetime $(M,g_{ab})$. As in~\cite{Kras04,Flanagan02} our convention is
that $u^au_a<0$ for timelike $u^a$. 
Let us examine the validity of the usual assumption near some
point $p\in M$. First, choose any `diamond neighbourhood' $D=
J^+(q)\cap J^{-}(r)$ of $p$ (where $q$ and $r$ lie in the causal past, resp.,
future of $p$) for which $(D,g_{ab}|_D)$, considered as a spacetime in
its own right, is globally conformal to the whole of Minkowski
space~\footnote{Such neighbourhoods always exist, as we may write the
metric in a neighbourhood of $p$ in the form $ds^2=e^{2\sigma}\,du\,dv$
for some smooth function $\sigma$ and coordinates $u,v$. We may assume that this neighbourhood
contains a diamond neighbourhood $D$ of $p$ corresponding to coordinate ranges
$|u|<u_0,|v|<v_0$, say, and by reparametrising $U=\tan(\pi u/(2u_0))$,
$V=\tan(\pi v/(2v_0))$, we see that $(D,g_{ab}|_D)$ is conformal to the
whole of Minkowski space.}. Now let $\gamma$ be any timelike
curve in $D$, parametrised by proper time $\tau$ within the range
$|\tau|<T$, say, and denote the two-velocity of $\gamma$ by $u^a$ and
its acceleration by $a^a =u^b\nabla_b u^a$. Because any state of the
field on $(M,g_{ab})$ induces a state of the field on $(D,g_{ab}|_D)$ with the
same renormalised stress-energy tensor~\footnote{The $n$-point functions
of the induced state are simply the restrictions to $D$ of the $n$-point
functions on $M$. The renormalised stress-energy tensor at any
point $p'\in D$ is obtained
from derivatives of the two-point function minus the Hadamard
parametrix; since the latter is determined by the local
geometry alone, and therefore independent of whether $p'$ is thought of 
as belonging to $M$ or $D$, it follows that the induced state has the
same renormalised stress-energy tensor as the original.}, we may apply Eq.~(1.7)
of~\cite{Flanagan02} to obtain
\begin{equation}
\int_\gamma \langle T_{ab}(\gamma(\tau))\rangle_\omega u^a u^b G(\tau) d\tau \ge
- \frac{1}{24\pi}\int_{-\infty}^\infty 
\left[\frac{G'(\tau)^2}{G(\tau)} +G(\tau) \left(a^a a_a + R\right)\right] d\tau
\label{eq:FQI}
\end{equation}
for any smooth nonnegative `sampling function' $G$ with compact support
in $(-T,T)$ and normalisation $\int_{-\infty}^\infty G(\tau)\,d\tau =1$~\footnote{The integrand on the
right-hand side of Eq.~\eqref{eq:FQI} is taken
to vanish outside the support of $G$. To convert Eq.~\eqref{eq:FQI} to a signature in
which $u^au_a>0$ for timelike $u^a$, it is necessary only to reverse the sign
of $a^aa_a$.}. This (absolute) QEI
is valid for any Hadamard state $\omega$ of the field on $M$.

We already see that the QEI bound consists of two parts: the flat
spacetime result, and correction terms due to the acceleration of the
curve and the scalar curvature of spacetime. As we now show,
the first part will dominate if $G$ is peaked on scales short in
comparison with those set by $R$ and $a^a$. Indeed, 
putting
\begin{equation}
A=\max\{0,\sup_\gamma a^a a_a\}\qquad{\rm and}\qquad
B=\max\{0,\sup_\gamma R\}\,,
\end{equation}
and replacing $G$ by the scaled version $G_{\tau_0}$ defined
by Eq.~\eqref{eq:scaled}, Eq.~\eqref{eq:FQI} implies that
\begin{equation}
\int_\gamma \langle T_{ab}(\gamma(\tau))\rangle_\omega u^a u^b G_{\tau_0}(\tau) d\tau \ge
-\frac{A+B}{24\pi}-\frac{C}{24\pi\tau_0^2}\,,
\label{eq:FQIs}
\end{equation}
where the constant $C$ is given in terms of the `unscaled' sampling function
$C =\int_{-\infty}^\infty G'(\tau)^2/G(\tau)\, d\tau$.
It is easy to find examples of $G$ supported within an interval of unit proper time
with $C$ of the order of $40$ (the minimum
value is $4\pi^2$~\footnote{Set $G(\tau)=2\sin^2(\tau\pi)$ for $0<\tau<1$ and
zero elsewhere, which attains $C=4\pi^2$. Although this function is only $C^1$, there exist smooth $G$ with
values of $C$ arbitrarily close to this value.}).
Accordingly, if $\tau_0\lesssim 10^{-3}\min\{A^{-1/2},B^{-1/2}\}$ and
$\tau_0<2T$ (i.e., sampling occurs within $D$) the second term in Eq.~\eqref{eq:FQIs}
dominates over the first by a factor of around $10$ and the flat space result may be safely
utilised, certainly for the order-of-magnitude considerations required in~\cite{FRworm,FPwarp}.
We have therefore justified the usual assumption. Three geometric
scales are relevant: the acceleration of the observer, the scalar curvature,
and the maximum size (as measured by $T$) 
of diamond neighbourhood globally conformal to the whole of Minkowski
space. The last of these becomes relevant when the spacetime contains
boundaries or singularities (cf.~\cite{FPR}).

Now let us turn to Krasnikov's claimed counterexample to the QEIs.
The essential content of~\cite{Kras04} is the following: smooth null
coordinates $(u,v)$ are chosen on the whole of 
two dimensional de Sitter space in such a way that the metric takes the
form
\begin{equation}
ds^2 = \frac{\alpha^2}{\sinh^2(u-v)}dudv
\end{equation}
in a region $W$ of the spacetime given by
\begin{equation}
W=\{(u,v): |u|,|v|<\frac{1}{2}|\log\tan\epsilon|,~u>v\}\,,
\end{equation}
where $\epsilon$ is a freely chosen parameter in the range $0<\epsilon\ll\pi/4$. These coordinates determine a
conformal map from de Sitter to a portion of Minkowski space (metric $ds^2=dudv$), and hence
a corresponding conformal vacuum state $\omega$ on de Sitter, whose stress-energy tensor may be computed
using Eq.~(6.134) of~\cite{BD} (adapted to our conventions). Expressing
the metric as $g_{ab}=e^{2\sigma}\eta_{ab}$, we have
[cf. Eq.~(2.5) in~\cite{Flanagan02}]
\begin{equation}
\langle T_{ab}\rangle_\omega =\frac{1}{12\pi}
\left[\partial_a\partial_b\sigma -
(\partial_a\sigma )(\partial_b\sigma)-
\eta_{ab}\eta^{cd}\partial_c\partial_d\sigma
+\frac{1}{2}\eta_{ab}\eta^{cd}(\partial_c\sigma
)(\partial_d\sigma)\right]\,,
\end{equation}
where $\partial_a$ denotes the covariant derivative for the
Minkowski metric $\eta_{ab}$. This may be evaluated most easily using the coordinates
$t=\frac{1}{2}(v-u)$ and $x=\frac{1}{2}(v+u)$, so that
$\sigma(t,x)=\log(-\alpha/\sinh(2t))$ on $W$. 
(Note that $\frac{1}{2}\log\tan\epsilon<t<0$ in this region.) 
The energy density measured by an observer moving along a curve of constant $x$
is then
\begin{equation}
\rho = e^{-2\sigma}\langle T_{tt}\rangle_\omega = -\frac{\cosh^2 (2t)}{6\pi\alpha^2}
\end{equation}
(Krasnikov has actually miscalculated this quantity~\footnote{The
equation given for $T_{uv}$ in~\cite{Kras04} contains two errors: it should read
$T_{uv}=RC/(96\pi)=1/(12\pi\sinh^2(u-v))$. In the process of 
lowering an index from Eq.~(6.136)
of~\cite{BD} and converting between different conventions for the
signature and definition of null coordinates, Krasnikov has lost a
factor of $-1/2$ in the first equality; a further algebraic error in the second
equality [note: $R=8/\alpha^2$] led him to the incorrect expression $T_{uv}=-1/(24\pi\sinh^2(u-v))$.}, but his argument
would equally apply to the corrected version). The exponential growth of this
quantity as $t$ becomes large and negative appears remarkable, but it 
is instructive to express it in
terms of the proper time parameter $\tau$, defined implicitly by
\begin{equation}
t(\tau) = \frac{1}{2}\log\tanh\frac{\tau}{\alpha}\,,
\end{equation}
and running over the range $\tau\in(\zeta,\infty)$ on $W$, where $\zeta=\alpha\tanh^{-1}(\tan\epsilon)$.
In terms of $\tau$, the energy density is
\begin{equation}
\rho=-\frac{\coth^2 (2\tau/\alpha)}{6\pi\alpha^2}\,,
\end{equation}
so $\rho\sim -1/(24\pi\tau^2)$ for small values of $\tau$, corresponding
to the intuitive understanding of QEIs. 

Krasnikov computes the {\em unweighted} integral of (minus) the energy density along a portion of
the curve $x=0$. Using the corrected form of the energy density, and
expressing everything in terms of proper time, this
gives
\begin{equation}
{\cal E} =\int_{\tau_0}^{\tau_1} -\rho(\tau) \,d\tau =\frac{\tau_1-\tau_0}{6\pi\alpha^2}
+\frac{1}{12\pi\alpha}
\left(
\coth\left(\frac{2\tau_0}{\alpha}\right)-\coth\left(\frac{2\tau_1}{\alpha}\right)
\right)
\end{equation}
from which one may see that ${\cal E}(\tau_1-\tau_0)$ \footnote{Note
that this quantity actually has dimensions
$(\hbox{energy})\times(\hbox{time})^2$.} can be made large ($O(\tau_0^{-1})$)
by making $\tau_0$ (and the cutoff $\epsilon$) small. Krasnikov asserts
that this violates the QEIs. But it does not: an average of the type
just made corresponds to a sampling function equal to the
characteristic function of the interval $(\tau_0,\tau_1)$, which is
nonsmooth and therefore outside the scope of QEIs. If one attempted to approximate the
characteristic function as a limit of smooth functions, the QEI bound would
diverge in the limit. So the QEI bound is effectively equal to $-\infty$, and there
is no contradiction with Krasnikov's result. 

This might give the impression that QEIs have nothing useful to say
about Krasnikov's example. Not at all: when the sampling function is
smooth, there is a finite bound, given by Eq.~\eqref{eq:FQI}. 
The fact that QEIs yield no useful information if the sampling
function is nonsmooth does not detract from the fact that they {\em do}
yield useful information when it is!

{}To remove all possible doubt,  
let us check that the energy density $\rho$ is indeed consistent with the QEI~\eqref{eq:FQI}. 
Curves with constant $x$ are geodesic, so $a^a=0$; we also have
$R=8/\alpha^2$. Writing $G(\tau)=g(\tau)^2$, and rearranging, we are
required to show that 
\begin{equation}
\int_{-\infty}^\infty \left[\alpha^2 g'(\tau)^2+(2-\coth^2(2\tau/\alpha)) g(\tau)^2 \right]\,d\tau \ge 0\,,
\end{equation}
for all real-valued smooth $g$ compactly supported in
$(\zeta,\infty)$. Since $\zeta$ can be made arbitrarily small by choice of
$\epsilon$, we must actually allow $g$ to have arbitrary compact
support in $(0,\infty)$. We may therefore discard the negative half of
the integration range. Integrating the first term by parts, our aim is now
to prove
\begin{equation}
\int_0^\infty g(\tau)\left\{-\alpha^2 g''(\tau)+[2-\coth^2(2\tau/\alpha)] g(\tau) \right\}\,d\tau \ge 0\,,
\end{equation}
or, equivalently, that $\ip{g}{Hg}\ge 0$ for all $g$, where $H$ is the operator
\begin{equation}
H= -\alpha^2 \frac{d^2}{d\tau^2} + 2-\coth^2\left(\frac{2\tau}{\alpha}\right) 
\end{equation}
and $\ip{\cdot}{\cdot}$ is the usual $L^2$-inner product on $(0,\infty)$
 (this approach was developed in detail in~\cite{FTii}).
It now suffices to observe that $H$ factorises as $H=A^*A$, with
\begin{equation}
A = \alpha\frac{d}{d\tau} - \coth\frac{2\tau}{\alpha}\,,
\end{equation}
so $\ip{g}{Hg}=\ip{g}{A^*Ag}=\ip{Ag}{Ag}=\|Ag\|^2\ge 0$ for all $g$ in our domain of
interest. We have therefore shown explictly that the QEI is
satisfied. A similar analysis would apply to any conformal vacuum. 

To summarise: we have shown that Krasnikov's supposed counterexample to
the QEIs is, in fact, entirely compatible with them. Moreover, we have given a
general proof, valid in two-dimensional curved spacetimes, to justify
the assumption made in~\cite{FRworm,FPwarp}, namely, that flat spacetime QEIs
adequately constrain curved spacetime behaviour for sampling times short in
comparison with natural geometric scales. 

\begin{acknowledgments}
I thank \'Eanna Flanagan and Tom Roman for useful comments.
\end{acknowledgments}


\begin{thebibliography}{ZZ}


\bibitem{Ford78} L.H. Ford,
                 Proc. R. Soc. Lond. A {\bf 364}, (1978) 227. 
\bibitem{Flanagan97} \'E.\'E. Flanagan, Phys. Rev. D {\bf 56}, (1997) 4922.
\bibitem{FRworm} L.H. Ford and T.A. Roman,
                Phys. Rev. D {\bf 53}, (1996) 5496.

\bibitem{FPwarp} M.J. Pfenning and L.H. Ford, 
                 Class. Quantum Grav. {\bf 14}, (1997) 1743.
\bibitem{Kras04} S. Krasnikov, `Counter example to the quantum inequality' {\tt
gr-qc/0409007v1}.

\bibitem{FordPfenning98} M.J. Pfenning and L.H. Ford,  Phys. Rev. D {\bf 57}, (1998) 3489.
\bibitem{FTeo}   C.J. Fewster and E. Teo,
                Phys. Rev. D{\bf 59}, (1999) 104016.
\bibitem{Flanagan02} \'E.\'E. Flanagan, Phys. Rev. D {\bf 66}, (2002) 104007.

\bibitem{AGWQI} C.J. Fewster,
                Class. Quantum Grav. {\bf 17}, (2000) 1897.
\bibitem{FVdirac} C.J. Fewster and R. Verch, Commun. Math. Phys. {\bf 225},
(2002) 331. 

\bibitem{Vollick00} D.N. Vollick, Phys. Rev. D {\bf 61}, (2000) 084022.

\bibitem{Pfenning_em} M.J. Pfenning, Phys. Rev. D {\bf 65}, (2002) 024009.

\bibitem{FewsterPfenning} C.J. Fewster and M.J. Pfenning, J. Math. Phys.
{\bf 44}, (2003) 4480.

\bibitem{FPR} L.H. Ford, M.J. Pfenning, and T.A. Roman,
Phys. Rev. D {\bf 57}, (1998) 4839. 

\bibitem{BD} N.D. Birrell and P.C.W. Davies, `Quantum fields in curved
space', (CUP, Cambridge, 1982).

\bibitem{FTii} C. J. Fewster and E. Teo, Phys. Rev D {\bf 61},
(2000) 084012.
\end{thebibliography}
\end{document}